\documentclass{epl}

\newcommand{\sech}{\mathrm {sech}}
\newcommand{\arctanh}{\mathrm {arctanh}}

\title{Enhanced Quantum Reflection of Matter-Wave Solitons}
\author{C. Lee\inst{1,2} \and J. Brand\inst{1}}

\institute{
  \inst{1} Max Planck Institute for the Physics of Complex
Systems, N{\"o}thnitzer Stra{\ss}e 38, 01187 Dresden,
Germany\\
  \inst{2} Nonlinear Physics Centre and ARC Centre of Excellence for
Quantum-Atom Optics, Research School of Physical Sciences and
Engineering, Australian National University, Canberra ACT 0200,
Australia }




\begin{document}

\maketitle

\begin{abstract}
Matter-wave bright solitons are predicted to reflect from a purely
attractive potential well although they are macroscopic objects
with classical particle-like properties. The non-classical
reflection occurs at small velocities and a pronounced switching
to almost perfect transmission above a critical velocity is found,
caused by nonlinear mean-field interactions. Full numerical
results from the nonlinear Schr\"{o}dinger equation are
complimented by a two-mode variational calculation to explain the
predicted effect, which can be used for velocity filtering of
solitons. The experimental realization with laser-induced
potentials or two-component Bose-Einstein condensates is
suggested.
\end{abstract}

In the framework of classical mechanics, a moving object will
never turn back until it reaches a turning point, at which the
radial velocity vanishes (the kinetic energy vanishes for one
dimensional systems). At the microscopic scale, where the wave
character of particles becomes important, quantum mechanics allows
the reflection of a particle in a classically allowed region even
when there is no classical turning point. {\em Quantum reflection}
can occur above a repulsive potential barrier or an attractive
potential well, and may take place in an attractive potential tail
\cite{Quantum-reflection} or at a potential step \cite{QR-Cote}. The
quantum reflection of cold atoms by a solid surface has triggered
great interest both for the fundamental understanding of the
implications of quantum mechanics and for potential applications
of mirrors in atom optics \cite{Single-Atom-Picture}. Recently,
Pasquini {\it et al.}\ reported the experimental observation of
the quantum reflection of atoms from a dilute Bose-Einstein
condensate (BEC) with up to 20 and 50\% efficiency \cite{QR-MIT}.

\begin{figure}
\onefigure[width=8.0cm,height=6.0cm]{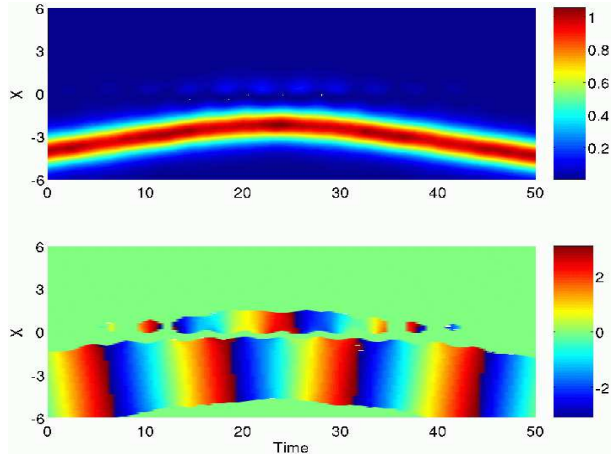}
\caption{\label{fig:carpet} Quantum reflection of a soliton
incident on an attractive potential well centred at $x=0$. The
initial conditions are $A=1$, $x_{0}=-4$ and $v=0.1$. The upper
panel shows the density and the lower one the phase evolution.}
\end{figure}

Matter-wave bright solitons are macroscopic quantum objects that
may act as classical particle-like objects maintaining their
integrity during collisions or while subjected to external forces.
They have been prepared as self-bound droplets of atomic BECs with
negative s-wave scattering lengths in quasi-one-dimensional
waveguides \cite{BEC-soliton,BEC-soliton2}. Previously we
considered conditions for the creation of soliton trains
\cite{carr1} and their propagation in harmonic traps \cite{Hai}.
In this Letter we show that a matter-wave soliton approaching an
attractive potential well may experience non-classical reflection.
In contrast to the finite probabilities of quantum reflection of
single atoms, the whole soliton reflects with very little
radiative loss as seen in Fig.~\ref{fig:carpet}, leading to a
significant enhancement of reflection due to nonlinear
interactions and macroscopic coherence. Above a critical velocity
we observe a sharp transition to almost complete transmission
while trapping is also possible in different parameter regimes.
%%%%%%%%%%%%%%%%%%%%%%%%%%%%%%%%%%%%%%%%%%%%%%%%%%%%%%%%%%%%%%%%
%%%%%%%%%%%%%%%%%%%%%%%%%%%%%%%%%%%%%%%%%%%%%%%%%%%%%%%%%%%%%%%%%

Solitons behave as classical particles according to perturbation
theory when they move in weak external potentials \cite{Kivshar}
or by virtue of Ehrenfest's theorem in semiclassical conditions
where the potential varies slowly on the size scale of the
soliton. Non-classical soliton scattering is expected when the
kinetic energy dominates over the soliton binding energy or in
specific resonant scattering scenarios and was previously
discussed for step-like \cite{earlier-works} and square potentials
\cite{Tamura} and impurities
\cite{Soliton-Impurity-Interaction,SII2,Flach}.

The effect of pronounced switching from quantum reflection to
transmission reported in this Letter fits into neither of these
categories as adiabatic conditions are broken by a strongly
localised potential well but nonlinearity dominates over kinetic
energy. Thus the soliton may remain intact. In the following we
will discuss the various parameter regimes in detail. We report
results of numerical simulations and study a two-mode variational
model, which gives valuable insights into the underlying
mechanism. We furthermore discuss possible experimental
realizations with optical dipole potentials and incoherent
solitons in two-component BECs and applications in velocity
filtering.

Consider a matter-wave soliton in a one-dimensional waveguide trap
approaching the centre of a localised attractive potential well
$V(x)$. The dynamics is described well by the Gross-Pitaevskii
equation (GPE), which we consider in the one-dimensional
approximation \cite{pitaevskiibook} and write in dimensionless
units as
\begin{equation} \label{eqn:GPE}
i\frac{\partial }{\partial t}\Psi
(x,t)=[-\frac{1}{2}\frac{\partial^{2}}{\partial x^{2}}%
-g \left| \Psi (x,t)\right| ^{2}+V(x)]\Psi (x,t).
\end{equation}
The parameter $g$ is proportional to the s-wave scattering length.
Without loss of generality, we assume that $g=0$ and $1$
correspond to the linear and nonlinear system, respectively.

For the linear system ($g=0$) with an attractive
sech-squared-shape potential $V(x)=-U{\sech}^{2}(\alpha x)$, known
as Rosen-Morse potential, the quantum mechanical reflection
probability $R$ is known exactly \cite{linear-systems},
\begin{equation} \label{eqn:reflprob}
R =\frac{\cos ^{2}[ \frac{\pi }{2}\sqrt{1+8U/\alpha ^{2}}]}{\sinh
^{2}(\pi v/\alpha )+\cos ^{2}[ \frac{\pi }{2}\sqrt{1+8U/\alpha
^{2}}]} ,
\end{equation}
and is determined by both the incident velocity $v$, which in our
units coincides with the particle momentum, and the system
parameters. Particles with large velocities are transmitted and
particles with small velocities reflected unless the cosine term
in the numerator of Eq.~(\ref{eqn:reflprob}) vanishes for specific
values of $U/\alpha^2$. In these special cases the transmission of
linear waves is reflectionless. Reflection is maximised when the
cosine term becomes unity. In this case, the reflection
probability $R$ is greater than 50\% for small velocities with $v<
v_R = 0.28  \alpha$.

For the nonlinear system ($g=1$) with the same potential, the
system supports travelling soliton solutions in the absence of the
external potential $V(x)$:
\begin{equation} \label{eqn:sol}
  \Psi=A\,{\sech} (A[x-x_{0} -v t]) e^{i(v x -\omega t)} ,
\end{equation}
where $v$ is the velocity, $A$ is the amplitude, and $x_0$ is the
initial position. The soliton frequency $\omega = v^2/2 +\mu$ can
be interpreted as the sum of the kinetic energy per particle and
the binding energy $\mu = -A^2/2$. We now consider the scattering
of solitons on the potential well $V(x)$ and find that the
situation may change dramatically compared to the linear case,
depending on the relevant scales of the system.

In order to avoid the classical Ehrenfest regime, the length scale
of the potential well $l_V = 1/ \alpha$ should be smaller than the
soliton length scale $l_S = 1/ A$ \cite{Kivshar}. A regime of linear wave
scattering can be expected when the time scale of interaction of
the soliton with a localised defect is shorter than a
characteristic time scale of the nonlinearity as suggested in
Ref.~\cite{Flach} for resonant interactions. For non-resonant
interactions we may rewrite this as a condition for the velocity,
which should be larger than the velocity $v_{\rm disp} = 4
|\mu|/l_S = 2 A^3$ characterising the dispersion of a linear wave packet
with the soliton parameters. In this regime we find a splitting of
the soliton into reflected and transmitted portions. Finally we
should compare the energy scale of the soliton $\omega$ with the
energy of the lowest bound state of the potential $V(x)$ in the
linear Schr\"odinger equation, which is \cite{linear-systems}
$E_0= - (\alpha^2/2) (\sqrt{2U/\alpha^2+1/4}-1/2)^2$. Following an
argument of Ref.~\cite{SII2}, we may expect a regime of strong
soliton-defect interaction if $\omega \leq E_{0}$ because an
overlapping between the incoming soliton and the localised bound
state may lead to a resonant population transfer from the soliton
into the bound state. In this regime, we find indeed that solitons
at small velocities are trapped by the potential well.

What are the possibilities to meet the required conditions by
choosing appropriate parameters for the soliton (\ref{eqn:sol}) and the
potential well? From the condition for the length scales for
non-Ehrenfest scattering $l_V < l_S$ we obtain $A < \alpha$. In order
to find a sharp transition between reflection and transmission, the
velocity scale for quantum reflection $v_R$ should be less than
$v_{\rm disp}$, which yields $0.28 \alpha < 3 A^3$. Both conditions
are compatible but require both $\alpha$ and $A$ to be larger than
roughly $1/3$. Finally, strong defect interactions can be avoided when
$\mu > E_0$. We find $A^2 < \alpha^2(\sqrt{2U/\alpha^2+1/4}-1/2)$,
which can always be satisfied if $U$ is chosen large enough.

%In order to maximise quantum reflection, we should avoid the
%regimes of Ehrenfest motion by choosing $l_V < l_S$
%, and the regimes of linear wave scattering and
%strong defect interactions by choosing $v<v_{\rm disp}$ and
%$\omega > E_{0}$, respectively. 
We now consider an initially well separated soliton with $A=1$
approaching a potential well with $U=4$ and $\alpha=2$.  Figure
\ref{fig:carpet} shows the result of a numerical integration of the
GPE (\ref{eqn:GPE}). The relevant scales are $l_{S}=1$, $l_{V}=1/2$,
$v_{\rm disp}=2$, $\mu=-1/2$, and $E_{0}=-4$.  The time evolution
shown in Fig.~\ref{fig:carpet} indicates that the matter-wave soliton
can indeed be reflected by the attractive potential. An important
observation from Fig.~\ref{fig:carpet} is that the potential well is
populated transiently by a small portion of the matter wave, with a
phase difference of $\pi$ compared to the soliton.
%%%%%%%%%%%%%%%%%%%%%%%%%%%%%%%%%%%%%%%%%%%%%%%%%%%%%%%%%%%%%%%%
\begin{figure}
\onefigure[width=8.0cm,height=6.0cm]{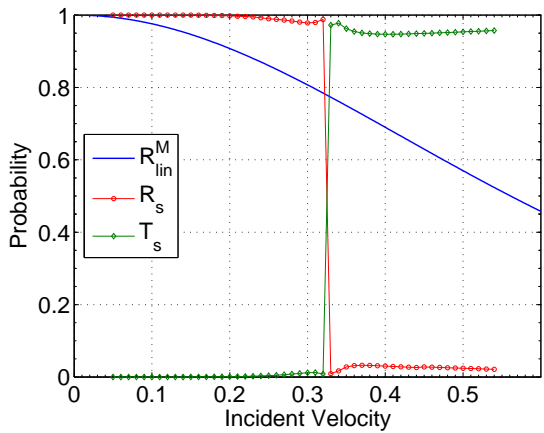}
\caption{\label{fig:RandT} Reflection $R$ and transmission $T$ versus
incident velocity $v$ for $U=4$ and $\alpha=2$. The $R_{lin}^{M}$ is
the maximum possible reflection in the linear system ($g=0$) obtained
by modifying the potential depth. The $R_{s}$ and $T_{s}$ corresponds
to the nonlinear system supporting an initial soliton with $A=1$
at $x_{0}=-10$.}
\end{figure}
%%%%%%%%%%%%%%%%%%%%%%%%%%%%%%%%%%%%%%%%%%%%%%%%%%%%%%%%%%%%%%%%%

Due to the mean-field interaction, the dependence of the
reflection probability on the incident velocity is dramatically
changed. In comparison with the smooth transition between
reflection and transmission in linear systems, the nonlinear
interaction makes the transition much sharper with a well-defined
critical velocity, seen in Fig.~2. Here, we calculated $R_{s}$ and
$T_{s}$ as the reflected and transmitted portion of atoms by
integrating the density $|\Psi(x,t)|^{2}$ after the collision at
$t=50$ on the left side and right side of the well, respectively.

In addition to reflected and transmitted waves, some atoms can be
trapped in the potential well. This process is particularly
important for small depths of the attractive potential well due to
resonant energy transfer \cite{SII2}, but can be mostly avoided if
the potential well is deep enough. Besides the travelling soliton
and the trapped mode, there are small amounts of radiation, in
particular when the incident velocity is in the critical region
between reflection and transmission. The soliton scattering
problem as as shown in the lower panel of Fig.~2 is dominated by
reflection when the incident velocity is below 0.32 but is
dominated by transmission when the incident velocity is above
0.33. In these two dominant regimes, at least $96\%$ of particles
are reflected or transmitted. We have varied the depth of the
potential well $U$ and found that the crossover velocity $v_R =
0.28 \alpha$ (obtained from $R_{lin}^{M}=1/2$) of the linear
system is a reasonable estimate for the critical velocity for the
quantum reflection of solitons. However, while the actual
reflection $R$ of Eq.~(\ref{eqn:reflprob}) for the linear system
shows a very sensitive dependence on $U$, we find no such strong
dependence for the scattering of solitons. The cases of
reflectionless scattering that occur in the linear problem for
specific values of $U=\alpha^2[(2N+1)^2-1]/8$ (where $N$ is a
non-negative integer) can be related to resonances at zero
scattering energy \cite{newton}. In the nonlinear case these
resonances do not survive because phase relationships are modified
by nonlinear interactions \cite{Flach}. For potentials with
reflectionless scattering $R(v)\equiv 0$ of linear waves, the
soliton still shows a reflection characteristic as in
Fig.~\ref{fig:RandT}, and thus presents an extreme enhancement of
the quantum reflection probability from exactly zero to more than
0.96.

The quantum reflection of the incident soliton can be interpreted
and analysed within a simple two-mode picture. To this end we
write
\begin{equation}
\Psi(x,t)=\Psi_{S}(x,t)+\Psi_{T}(x,t) ,
\end{equation}
where  $\Psi _{S}(x,t)$ represents the soliton and $\Psi
_{T}(x,t)$ is an immobile trapped mode which is of importance if
the potential is sufficiently deep and localised to escape the
perturbative regime. When the two modes overlap, the off-diagonal
terms of the attractive potential, $\Psi_{S}^{*}(x,t) V(x)
\Psi_{T}(x,t)$ and $\Psi_{T}^{*}(x,t) V(x) \Psi_{S}(x,t)$, will
result in a repulsive force between two modes when they are out of
phase. Additionally, similar to the interaction between two
coherent solitons \cite{Coherent-soliton-interaction}, the
mean-field interaction will induce a repulsive (or attractive)
force between two modes when they are out of phase (or in phase).
When the soliton moves close to the potential centre, the soliton
will be reflected if the repulsion overcomes the attractive
effects.

A collective coordinate approach  has been used in
Ref.~\cite{SII2} to study the interactions of the soliton and the
defect mode for the case of a delta function defect potential
$V_D(x) = -A_P \delta(x)$. The delta defect with $A_{P} \sim
2U/\alpha$ is a reasonable approximation for a strongly localised
potential well as long as $l_V \ll l_S$ and thus we may expect to
obtain some insight from the model of Ref.~\cite{SII2} for our
case. The variational ansatz for the trapped mode is the
parametrised solution of the nonlinear localised state with the
potential $V_D(x)$ and is written as $\Psi_{T}=A_{T}\sech
[A_{T}|x|+\arctanh (\frac{A_P}{ A_{T}})]\exp(i\phi _{T})$. Here,
the trapped mode amplitude $A_{T}$ and phase $\phi_{T}$ appear as
time-dependent variational parameters.  For the soliton mode we
choose $\Psi_S=A_S \sech [A_S x-Q]\exp (i V x+i\phi _{S})$, where
the position $Q$, velocity $V$, amplitude $A_S$, and phase $\phi
_{S}$ are variational parameters. The Lagrangian for the
collective coordinates is derived from the standard variational
formulation of the GPE (\ref{eqn:GPE}) \cite{pitaevskiibook} and
simplified by neglecting interaction terms not proportional to
$V(x)$ \cite{SII2} to yield
\begin{equation} \label{eqn:Lagrangian}
\begin{array}{rcl}
\nonumber L &=& -2A_{S}\frac{d \phi_{S}}{dt}-2Q\frac{dV}{dt}
-2(A_{T}-A_{P})\frac{d
\phi_{T}}{dt}+\frac{1}{3}A_{S}^{3}-A_{S}V^{2}+\frac{1}{3}A_{T}^{3}-
\frac{2}{3}A_{P}^{3}\\
&&
+A_{P}A_{S}^{2}{\sech}^{2}Q+2A_{P}A_{S}\sqrt{A_{T}^{2}-A_{P}^{2}}\,{\sech}Q
\, \cos (\phi _{T}-\phi _{S}).
\end{array}
\end{equation}
The equations of motions for the  collective coordinates $Q$, $V$,
$A_S$,  $\phi _{S}$, $A_{T}$, and $\phi_{T}$ are given by the
Euler-Lagrange equations of $L$ and can be solved
straightforwardly by numerical integration.
%%%%%%%%%%%%%%%%%%%%%%%%%%%%%%%%%%%%%%%%%%%%%%%%%%%%%%%%%%%%%%%%
\begin{figure}
\onefigure[width=8.0cm,height=6.0cm]{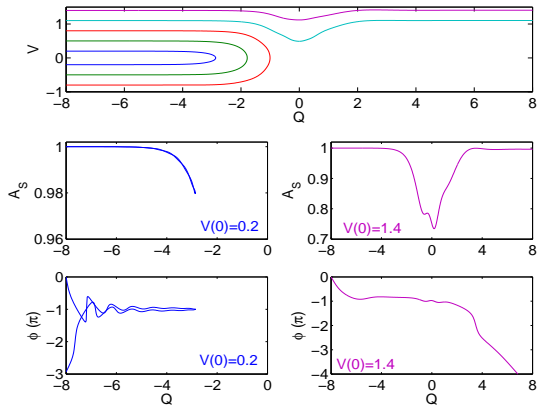}
\caption{Soliton trajectories from the variational
Eq.~(\ref{eqn:Lagrangian}) are shown for $A_{P}=2.0$ and initial
conditions $A_{S}(0)=1.0$ and $Q(0)=-8.0$. The upper panel shows
the velocity $V$ and position $Q$ for different incident
velocities $V(0)$. Below, the amplitude $A_S$ and relative phase
$\phi=\phi _{T}-\phi _{S}$ are shown for the particular
trajectories with $V(0) = 0.2$ (reflected, left panels) and $V(0)
= 1.4$ (transmitted, right panels).}
\end{figure}
%%%%%%%%%%%%%%%%%%%%%%%%%%%%%%%%%%%%%%%%%%%%%%%%%%%%%%%%%%%%%%%%%

The numerical results as seen in Fig.~3 show that the incident
soliton can indeed be reflected by the attractive potential well
when the incident velocity is sufficiently small. Otherwise, when
the incident velocity is larger than a certain critical value, the
soliton will travel through the potential. When the distance
between the soliton and the trapped mode is small enough,
population transfer between them may occur. For this reason, the
soliton mode obtains a time delay when it approaches the potential
centre and is decelerated. The deceleration of the soliton mode
becomes more and more significant with decreasing incident
velocity. For sufficiently slow incident velocity, the velocity
can finally change direction, and the incident soliton is
reflected. This behaviour is obviously contrary to the classical
case. To find where this counterintuitive behaviour comes from, we
track the evolution of the relative phase between the two modes.
We find that the two modes are out of phase (the relative phase
oscillates around $\pi$) when the soliton mode approaches the
potential centre. Thus the overlapping between modes induces a
repulsive force between them. Clearly, these variational results
are qualitatively consistent with the directly numerical results
from the GPE (1). However, differences appear in quantitative
results such as the critical velocity between reflection and
transmission.

The sharp transition between reflection and transmission may be
useful for filtering solitons according to their velocity. As a
particular potential has an associated critical velocity $v_c$, it
may discriminate  between solitons with velocities below and above
$v_c$. A sequence of potentials with different critical velocities
may be used to specifically select solitons with velocities in a
given interval. In conjunction with a potential ramp where
solitons are accelerated with $\ddot{Q} = -M V'(Q)$ it should also
be possible to select solitons according to their mass $M$.

The specific choice of the scattering potential $V(x)$ as a
Rosen-Morse potential is not essential for the observed effects and
was made for convenient comparison of the soliton scattering to the
analytically known reflection probability of the linear system. We
have varied the shape of the potential well in numerical simulations
in order to study its influences on the nonlinear scattering
properties. In the case of a Gaussian potential well, the picture
remains qualitatively the same with very little variation compared to
a Rosen-Morse potential with the same width and depth. In the case of
rectangular potential shape the situation changes and instead of a
sharp transition between almost complete reflection and transmission
we find an interval of velocities where the transition occurs. Within
the transition interval the soliton is typically split into two wave
packets which propagate as solitons in addition to some
small-amplitude radiation. A qualitative explanation may be that wave
diffraction on the sharp edges of the well effectively reduces the
coherence required for cohesion of the soliton. The details of these
studies will be published elsewhere \cite{note}.

% Generally, for other non-special potential parameters, there exist
% a transition region from full reflection to full transmission of
% matter-wave solitons, and its width depends on those parameters.
% That is, there are two critical velocities $V_{1}^{C}<V_{2}^{C}$,
% and full reflection (or transmission) occurs when the incident
% velocity $v_{i}<V_{1}^{C}$ (or $v_{i}>V_{2}^{C}$). In the
% transition region, $V_{1}^{C}<v_{i}<V_{2}^{C}$, reflection,
% transmission and trapping coexist. Nevertheless, due to the
% nonlinear mean-field interaction, the transition from full
% reflection to full transmission is still more quickly then the
% linear counterparts. Similar enhanced behaviours also exist in
% other types of attractive potentials, such as, Gaussian and
% rectangular potentials. The details will be published elsewhere in
% future \cite{note}.

In order to observe the quantum reflection of solitons induced by
a finite potential well, one can use a tightly focussed
red-detuned laser beam to form an attractive Gaussian shape
potential. In order to see quantum reflection, the laser focus has
to be brought down to the order of the soliton length scale, which
was $l_S \approx 1.7\mu$m in the experiment of
Ref.~\cite{BEC-soliton2}. A different possibility could be
realized by another, heavy and incoherent soliton. Such could be
realized with a two-component atomic BEC \cite{Two-component-BEC}
with all negative $s$-wave lengths, which can be adjusted by using
Feshbach resonances \cite{FR2}. In a quasi one-dimensional
external trap, the system supports well-separated incoherent
solitons. When the difference between the soliton amplitudes
controlled by the atomic numbers is very large, the soliton with
large atomic number is almost unchanged when they approach each
other. In the limit of only one atom in the small component, the
intra-component mean-field interaction in the small component is
absent and the system is equivalent to the linear case  of Eq.
(1). Otherwise, we have a wider soliton with small atom number
moving in an attractive potential well of Rosen-Morse type formed by the
narrower soliton with a large number of atoms. For this case, our
numerical results show that quantum reflection of the wider
soliton does occur for sufficiently slow relative velocity
\cite{note}.

In summary, we predict that quantum reflection can occur to a kind
of macroscopic quantum objects, atomic matter-wave bright
solitons. The pronounced switching between reflection and
transmission is a characteristic behaviour that should be
observable for sufficiently well localised and deep potential
wells, such as created by a strongly focussed red-detuned laser
beam or a second, incoherent soliton. Possible applications lie in
sensitive velocity and mass filtering for nonlinear atom optics
but may extend into other areas like  soliton-based optical
computing. The enhanced switching behaviour is induced by the nonlinear
mean-field interaction and will also occur in the interaction with
other types of potentials. Interesting further questions like the
possibility of quantum reflection of matter-wave solitons from solid
surfaces lie outside the scope of 
this Letter.

\acknowledgments We acknowledge useful comments by \Name{S. Flach}
and \Name{Y. S. Kivshar}. This work is partially supported by the
Australian Research Council (ARC).

\end{document}